\documentclass[usegraphicx]{mn2e}

\title[Circumstellar discs in the $\eta$ Cha cluster]
{Infrared study of the $\eta$ Chamaeleontis cluster and the
longevity of circumstellar discs}
\author[A. R. Lyo et al.]
{A-Ran Lyo,$^{1}$\thanks{E-mail: arl@ph.adfa.edu.au (ARL); 
wal@ph.adfa.edu.au (WAL); eem@as.arizona.edu (EEM); 
edf@astro.psu.edu (EDF); ecsung@mso.anu.edu.au (ECS); 
lcrause@artemisia.ast.uct.ac.za (LAC)}
Warrick A. Lawson,$^{1\star}$ 
Eric E. Mamajek,$^{2\star}$ 
Eric D. Feigelson,$^{3\star}$
\newauthor
Eon-Chang Sung$^{4, 5\star}$ and 
Lisa A. Crause$^{6\star}$\\ 
$^{1}$School of Physics, University of New South Wales, 
Australian Defence Force Academy, Canberra ACT 2600, Australia\\
$^{2}$Steward Observatory, University of Arizona, 933 N Cherry
Avenue, Tuscon AZ 85721, USA\\
$^{3}$Department of Astronomy and Astrophysics, Pennsylvania 
State University, University Park PA 16802, USA\\
$^{4}$Research School of Astronomy \& Astrophysics, Australian 
National University, Canberra ACT 2600, Australia\\ 
$^{5}$Korea Astronomy Observatory, Taejon 305-348, Korea\\
$^{6}$Department of Astronomy, University of Cape Town,
Private Bag, Rondebosch 7700, South Africa}

\begin{document}

\date{Accepted .................... Received ....................}

\pagerange{\pageref{firstpage}--\pageref{lastpage}}
\pubyear{2002}

\maketitle

\label{firstpage}

\begin{abstract}
We have analyzed {\it JHKL\,} observations of the stellar
population of the $\approx 9$ Myr-old $\eta$ Chamaeleontis cluster.
Using infrared (IR) colour-colour and colour-excess diagrams,
we find the fraction of stellar systems with near-IR excess 
emission is $0.60 \pm 0.13$ (2$\sigma$).  This results implies
considerably longer disc lifetimes than found in some recent 
studies of other young stellar clusters.   For the classical 
T Tauri (CTT) and weak-lined T Tauri (WTT) star population, 
we also find a strong correlation between the IR excess and 
H$\alpha$ emission.  The IR excesses of these stars indicate 
a wide range of star-disc activity; from a CTT star showing 
high levels of accretion, to CTT -- WTT transition objects with
evidence for some on-going accretion, and WTT stars with weak or
absent IR excesses.  Of the 15 known cluster members, 4 stars 
with IR excesses $\Delta(K-L) > 0.4$ mag are likely experiencing 
on-going accretion owing to strong or variable optical emission.
The resulting accretion fraction ($0.27 \pm 0.13$; 2$\sigma$) 
shows that the accretion phase, in addition to the discs
themselves, can endure for at least $\sim 10$ Myr.

\end{abstract}


\section{Introduction}

Circumstellar discs are a natural by-product of the star formation 
process (e.g. Shu, Adams \& Lizano 1987). Lada et al. (2000) found 
that 97 percent of the optical proto-planetary discs in the 
Trapezium cluster exhibit excess in the {\it JHKL\,} colour-colour 
diagram, indicating that the most likely origin of the observed IR 
excesses are the circumstellar discs.  The {\it disc lifetime\,} 
derived by examining the fraction of IR excess stars 
(the {\it disc fraction\,}) of young stellar populations as a 
function of age provides an empirical limit on the duration of 
the disc accretion phase.  This is critical for understanding the 
evolutionary paths followed by PMS stars in the HR diagram, their 
angular momentum histories, and the timescales available for planet 
building (e.g. Hartmann et al. 1998; Hillenbrand et al. 1998; 
Telesco et al. 2000). 

Considerable debate has waged concerning the longevity of  discs around
low mass PMS stars.  The effort is difficult because samples are often
incomplete and biased:  CTT stars are typically located by H$\alpha$
and near-IR excess surveys, while WTT stars are mostly found
through X-ray surveys (Feigelson \& Montmerle 1999).  An influential
early study on the timescale for disc dissipation by Strom et al. 
(1989), based on several dozen PMS stars in the Taurus-Auriga complex, 
reported the CTT/WTT transition occurs around an age $t \approx 3$ Myr.
This result is supported by a recent {\it JHKL\,} survey of 7 clusters 
with mean ages from 0.5 to 5 Myr that shows half the stars lose their
discs within 3 Myr and essentially all lose their discs in 6 Myr
(Haisch, Lada \& Lada 2001).  At a later age of $t \simeq 13$ Myr,
only 1/110 Sco-Cen late-type stars show spectroscopic CTT emission
lines and $K$-band excesses (Mamajek, Meyer \& Liebert 2002).

However, other studies suggest discs are more enduring.  No evolution
in disc fraction is found in the stellar populations of the Orion 
Nebula Cluster from $t < 0.1-2$ Myr (Hillenbrand et al. 1998), and
in NGC 2264 from $t < 0.1-5$ Myr (Rebull et al. 2002). The Chamaeleon
I cloud population, where the sample is enhanced through X-ray and
{\it ISO\,} surveys, shows no difference in the age distribution of 
CTT and WTT stars from $t < 1-10$ Myr (Lawson, Feigelson \& Huenemoerder 
1996).  Spectroscopic study of the $\sim 10$ Myr-old TW Hydrae 
Association members found active accretion in 2 stars; TW Hya and 
Hen 3-600A (Muzerolle et al. 2000), albeit at mass accretion rates 
$1-2$ magnitudes lower than is derived for younger accreting systems.

The measurement of disc lifetimes is muddied by several other issues.
First, despite the broad overlap in CTT and WTT age distributions, few
intermediate cases are known suggesting that the transition between 
the two states is rapid (Wolk \& Walter 1996).  Secondly, some studies 
suggest that two disc lifetimes must be considered; one associated with 
a rapid decline in accretion onto the star and another associated with 
a slower dissipation of the outer disc (Clarke, Gendrin \& Sotomayor 
2001).  Third, efforts to reconcile rotational distributions of both 
PMS and zero-age main sequence stellar clusters assuming star-disc 
rotational coupling during the accretion phase have deduced relatively 
long $10-15$ Myr disc lifetimes for slowly rotating stars (Bouvier, 
Forestini \& Allain 1997; Tinker, Pinsonneault \& Terndrup 2002), but 
$\sim 3$ Myr disc lifetimes may suffice if the stellar interiors have 
sufficient radial differential rotation (Barnes, Sofia \& Pinsonneault 
2001).  Finally, the astrophysical processes leading to disc dissipation 
are highly uncertain.  Possible mechanisms include accretion onto the 
star, bipolar outflow, stellar winds, photoevaporation, close gravitational 
encounters and incorporation of disc material into planets (see the 
review by Hollenbach, Yorke \& Johnstone 2000).

The recently discovered $\eta$ Chamaeleontis cluster has the potential
of clarifying some of the observational issues.  It is a nearby ($d
\simeq 97$ pc),  compact (extent $\sim$1 pc) and coeval ($t = 9 \pm 1$
Myr; Lawson \& Feigelson 2001) system of  PMS stars with a small 
(15 known primaries) population
of stars spanning a relatively large range in mass ($M = 0.2-3.4$
M$_{\odot}$; Mamajek, Lawson \& Feigelson 1999, 2000;  Lawson et
al.\ 2001, 2002).  The cluster includes $\eta$ Cha (spectral type
B8), HD 75505 (A5), the A7+A8 binary and $\delta$ Scuti system RS Cha,
11 WTT stars (K5$-$M4) and 1 CTT star (M2). Importantly, this census is
known to be virtually complete in the inner region from a combination
of a deep $ROSAT$ exposure, optical photometry and proper motion study
of the field. Unlike most other PMS populations, these membership
criteria are independent of the presence or absence of  circumstellar
discs. We can thus use the $\eta$ Cha population as an unbiased
laboratory to investigate the fraction of IR excess and accretion
at a critical intermediate-age phase of disc evolution.

\section{Observations and Data Reduction}

A $L$-band (3.5 $\mu$m) map of the $\eta$ Cha cluster was made 
during 1999 July $26-28$ with the 0.6-m South Pole Infrared 
Explorer (SPIREX) telescope, using the Abu camera which had a 
1k $\times$ 1k InSb detector array. The Abu camera had a plate 
scale of 0.6 arcsec pixel$^{-1}$, giving a field-of-view (FOV) 
of 100 arcmin$^{2}$.  Our observations were conducted in good 
conditions with below-average background levels.  The raw data 
frames were pipeline-reduced at the Rochester Institute 
of Technology before being delivered to us.
Using custom {\tt IRAF} routines written for the SPIREX/Abu 
system, we then merged the multiple dithered images made of 
each field into single frames for further analysis.

{\it JHK$_{N}$}-band images of the cluster members were obtained 
during 2002 March $2-5$ with the Cryogenic Array Spectrometer/Imager 
(CASPIR) on the 2.3-m telescope operated by Mount Stromlo and Siding
Spring Observatories (MSSSO).  CASPIR uses a 256 $\times$ 256 InSb 
detector array and we selected a FOV of 4.5 arcmin$^{2}$ giving a 
resolution of 0.5 arcsec pixel$^{-1}$.  Our observations were 
obtained under photometric conditions in $< 2$ arcsec seeing.  
The CASPIR images were linearized, sky-subtracted and flat-fielded 
using customized {\tt IRAF} routines based on, e.g. {\tt ccdproc}.

The SPIREX and CASPIR images were analysed using photometric 
routines (such as {\tt phot}) running within {\tt IRAF}. Fluxes 
were extracted and calibrated by comparison to standard stars
listed on the SPIREX homepage\footnote{See the SPIREX homepage 
at {\tt http://pipe.cis.rit.edu/}.} and IRIS Photometric standards 
(Carter \& Meadows 1995), respectively. 

Examination of the image profiles for the early-type cluster
members ($\eta$ Cha, RS Cha and HD 75505) suggested the onset 
of saturation in the CASPIR $J$ and $H$ frames.  For this reason 
we obtained {\it JHKL\,} photometry (in the SAAO system;
Carter 1995) for these 3 stars ({\it JHK\,} only for HD 75505)
using the 0.75-m telescope and Mark II IR photometer at the 
South African Astronomical Observatory (SAAO).  These 
data, along with measurements of IR standard stars, were 
obtained during the week of 2002 April 30 to May 6.  The SAAO 
{\it KL\,} data shows close agreement ($\pm 0.03$ mag) to the 
CASPIR $K$ and SPIREX $L$ data except for the $L$-band 
measurement of RS Cha, which we discuss below.

Table 1 lists the {\it JHKL\,} photometry of the $\eta$ Cha 
cluster members.  We adopt 1$\sigma$ uncertainties of 0.03 mag
for the SAAO {\it JHK\,} and the CASPIR {\it JHK$_{N}$\,} data.
For the brighter $L$-band sources ($L < 9$) we adopt a 1$\sigma$ 
uncertainty of 0.05 mag.  Fainter $L$-band magnitudes are 
uncertain by 0.1 mag. For several RECX stars we can compare our 
observations to on-line DENIS {\it JK\,} observations and to 
{\it JHKL\,} data published by Alcal\'a et al. (1995).  In most
cases, the magnitudes differ by $< 0.1$ mag, which we consider 
to be insignificant given differences between IR photometric 
systems, and the intrinsic variability of these stars 
(Lawson et al. 2001).  A special case is the eclipsing 
binary RS Cha.  Using the ephemeris of Clausen \& Nordstr\"om
(1978) we find the SPIREX $L$-band measurement was obtained 
during the secondary eclipse, whereas the CASPIR and SAAO 
magnitudes were obtained at maximum light.

In the following sections we make use of the SAAO photometry
for the 3 early-type stars; otherwise we adopt our CASPIR and
SPIREX observations.  Colours derived from the individual
magnitudes were transformed to the `homogenized' IR system of
Bessell \& Brett (1988) making use of equations provided
by Bessell \& Brett (1988) and McGregor (1994, 1997).  For 
($J-H$) and ($H-K$) colours derived from the CASPIR photometry, 
the correction is small ($< 0.02$ mag).  No correction was 
applied to the (CASPIR $K$ -- SPIREX $L$) colour due to the 
mix of photometric systems, and any correction is likely 
swamped by the $0.05-0.1$ mag uncertainty in the $L$-band 
photometry.  For colours derived for the early-type stars 
from SAAO photometry, the effect of transforming the colours 
from the SAAO system to the `homogenized' system is a correction 
of $\approx -0.02$ mag.

\begin{table}
\centering
\caption{{\it JHKL\,} photometry for members of the $\eta$ Cha cluster.}
\begin{tabular}{@{}lrrrr@{}}
\hline
Star & $J$~ & $H$~ & $K$~ & $L$~ \\
\hline\\
\multicolumn{5}{l}{\hspace*{-2mm}CASPIR/SPIREX observations}\\
\\
RECX 1             &  8.20 &  7.60 &  7.27 &  6.97 \\
RECX 3             & 10.57 &  9.86 &  9.61 &  9.11 \\
RECX 4             &  9.69 &  8.92 &  8.66 &  8.32 \\
RECX 5             & 10.99 & 10.29 &  9.96 &  9.26 \\
RECX 6             & 10.42 &  9.74 &  9.46 &  8.93 \\
RECX 7             &  8.61 &  7.92 &  7.69 &  7.48 \\
RECX 9             & 10.53 &  9.83 &  9.50 &  8.82 \\
RECX 10            &  9.68 &  8.95 &  8.78 &  8.48 \\
RECX 11            &  8.85 &  8.06 &  7.71 &  7.08 \\
RECX 12            &  9.38 &  8.70 &  8.51 &  8.02 \\
$\eta$ Cha         &  ---  &  ---  &  5.73 &  5.62 \\
RS Cha             &  ---  &  ---  &  5.45 &  5.84 \\
HD 75505           &  ---  &  ---  &  6.98 &  6.81 \\
ECHA J0841.5--7853 & 11.90 & 11.30 & 10.94 &  ---  \\
ECHA J0843.3--7905 & 10.66 &  9.93 &  9.45 &  8.40 \\
\\
\multicolumn{5}{l}{\hspace*{-2mm}SAAO observations}\\
\\
$\eta$ Cha         &  5.68 &  5.72 &  5.75 &  5.65 \\
RS Cha             &  5.60 &  5.46 &  5.43 &  5.35 \\
HD 75505           &  7.10 &  7.02 &  7.00 &  ---  \\
\hline
\end{tabular}
\end{table}

\section{Analysis of results}

\subsection{The infrared colour excess as a disc indicator}

Colour-colour diagrams, constructed from multi-wavelength IR
photometric and imaging surveys, have been shown to be a powerful 
tool for identifying IR excesses and circumstellar discs around 
stars in young clusters and star-forming regions.
In particular, $L$-band data combined with shorter 
wavelength observations permits evaluation of the fraction of 
sources with IR excess emission from circumstellar discs -- the
{\it disc fraction\,} (see, e.g. Haisch, Lada \& Lada 2000; Kenyon 
\& G\'omez 2001; Lada et al. 2000).  As Kenyon \& G\'omez (2001)
convincingly showed in their SPIREX $L$-band study of the Cha I
molecular cloud, $L$-band photometry is nearly essential for a
meaningful evaluation of the disc fraction in a young stellar 
population.  {\it JHK\,} observations alone do not extend to
a long enough wavelength range to enable a complete or unambiguous 
census of circumstellar discs in young clusters.  Data obtained
at 3.5 $\mu$m provides more contrast relative to photospheric 
emission from the central star compared to the shorter wavelength 
observations.

Figure 1 shows (a) {\it JHK\,} and (b) {\it JHKL\,} colour-colour 
diagrams of the $\eta$ Cha cluster members.  All 15 cluster members 
are shown in Figure 1(a).  Only 14 are shown in Figure 1(b); the
M4 cluster member ECHA J0841.5-7853 was not observed at $L$ band. 
In these diagrams the solid curves are the locus of colours 
corresponding to main-sequence stars of spectral types B8 -- M5 
(Bessell \& Brett 1988), which encompasses the range of spectral 
types of the cluster members.  In each figure, the dashed parallel 
lines define the reddening bands derived from relationships given 
by Bessell \& Brett (1988), where $E(J-H)$/$E(H-K) = 1.95$ 
and $E(J-H)$/$E(K-L) = 2.47$, respectively.  

\setcounter{figure}{0}
\begin{figure*}
\begin{center}
\includegraphics[height=85mm]{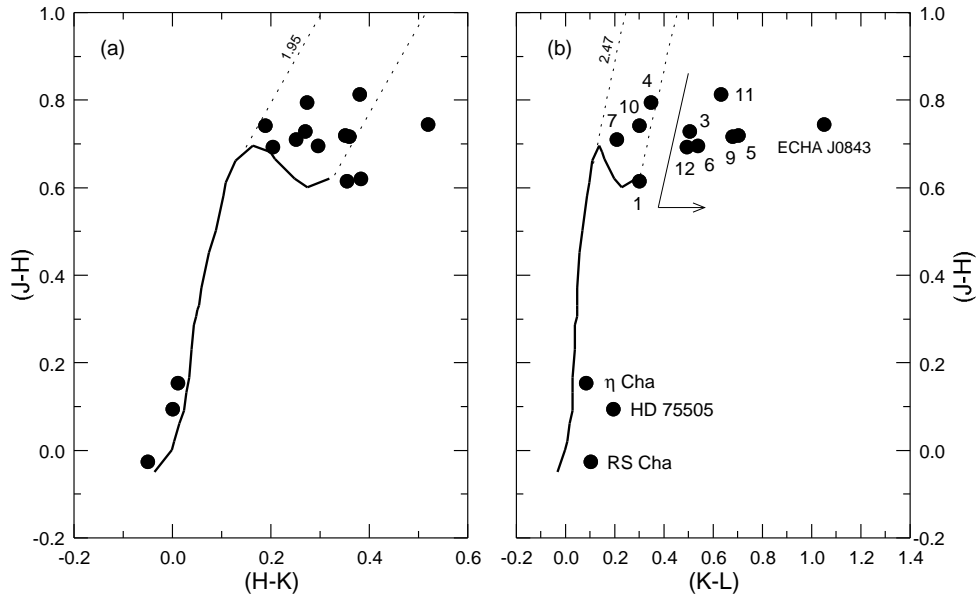}
\caption{(a) ($J-H$)/($H-K$) and (b) ($J-H$)/($K-L$) 
colour-colour diagrams for the $\eta$ Cha cluster.  
The bold lines are the locus of main-sequence stars from spectral 
type B8 -- M5, and the dashed lines represent the reddening band 
with gradient derived from equations given by Bessell \& Brett (1988).
In (b) the late-type stars are identified by their RECX number
(Mamajek et al. 1999), except for the CTT star ECHA J0843.3--7905
(Lawson et al. 2002).  The vector delineates late-type stars 
likely to have an IR excess; see Section 3.1.}
\end{center}
\end{figure*}

Stars which lie in the right of the reddening band, after due
consideration of uncertainties in the reddening law and in the 
photometry, are IR excess objects and therefore circumstellar 
disc candidates.  As Kenyon \& G\'omez (2001) found in their 
study of Cha I, only those stars with the largest IR excesses 
fall to the right of the reddening band in the {\it JHK\,} 
plane, whereas many more stars fall to the right of the 
reddening band once $L$-band data is available.
If photometric errors were negligible, we could count all 
late-type stars with ($K-L$) immediately to the right of the 
reddening band; however with $0.05-0.1$ mag uncertainties in 
the $L$-band data, we count only those late-type stars with 
($K-L$) colours 0.1 mag redder than the reddening band 
as IR-excess objects.  This criterion also largely eliminates 
uncertainty in the reddening law as a contributor to the disc 
fraction, as the reddening line is almost constant in ($K-L$)
colour over the narrow range of ($J-H$) colours for the 
late-type stars.

From Figure 1(b), the `gap' in the ($K-L$) colours between
0.34 and 0.49 allows us to count 7 stars as the most likely 
number of IR excess objects in the low-mass 
population\footnote{This criterion is similar to that adopted 
by Kenyon \& Hartmann (1995) for PMS stars in the Taurus-Auriga 
molecular cloud, where ($K-L$) = 0.4 delineated the disced 
and the discless populations.}.  
To estimate the uncertainty in this number, we 
calculated the variation in the number of stars if the photometric 
uncertainties are considered at the 2$\sigma$ level, i.e. twice 
the level of the adopted photometric errors (see Section 2).  
Now we find $7 \pm 2$ could be counted as IR excess objects.  

Use of standard IR colour-colour diagrams such as Figure 
1(b) could lead to an under-estimate of the disc fraction 
since K-type stars need to have a colour excess $\Delta(K-L) 
> 0.3$ to be counted, whereas late-M stars need only 
$\Delta(K-L) > 0.1$.  An alternative IR colour-colour 
diagram which largely eliminates this problem is shown in 
Figure 2.  In this plane the reddening band is narrow, and 
the reddening vector is parallel to the main-sequence, 
enabling stars to be counted using the ($K-L$) excess as 
the criterion.  For $\Delta(K-L) > 0.25$ above the 
main-sequence line of Bessell \& Brett (1988), we count 
the same 7 stars as the most-likely IR excess objects.  
If we adopt the same 2$\sigma$ error budget as above, 
we again count $7 \pm 2$ as the range of IR excess stars.
Also, as the faint cluster member ECHA J0841.5--7853 has 
{\it JHK\,} colours consistent with its M4 spectral type, 
it is unlikely to have a large IR excess, and so we conclude
$7 \pm 2$ out of the 12 known late-type members are IR
excess objects.

For the 3 early type stars, we may be justified in defining
a lower colour excess as being significant.  $\eta$ Cha is a
$\beta$ Pic-disc source, and with 0.05 mag 
uncertainty in the SAAO $L$-band flux the star has a ($K-L$) 
excess that is significant at the 3$\sigma$ level. For HD 75505, 
a 0.05 mag uncertainty in the SPIREX $L$-band magnitude gives
rise to a 4$\sigma$ excess. Thus with $> 3\sigma$ certainty,
2 of the 3 early-type cluster members are IR excess objects.  

\subsection{Comparison with other accretion indicators -- 
H$\alpha$ equivalent widths}

Since an IR excess most likely indicates the presence of a 
circumstellar 
disc, we might then expect to find a correlation between the 
IR excess and accretion indicators such as enhanced H$\alpha$ 
emission, which is believed to originate in magnetospheric columns 
allowing transport of disc material to the stellar surface. Two 
recent studies have demonstrated a correlation between 
the H$\alpha$ equivalent width ($EW$) and the ($K-L$) colour in 
Cha I PMS stars (Kenyon \& G\'omez 2001), or the IR colour excess 
in the NGC 2264 population (Rebull et al. 2002). Comparison 
between the H$\alpha$ $EW$ and the {\it colour excess\,} is 
desirable to eliminate redundancy caused by the spread in 
{\it colour\,} across a stellar population.

Determination of the colour excess for individual stars requires 
knowledge of the spectral type.  For $\eta$ Cha cluster members, 
spectral types have been determined from high resolution studies 
for several RECX stars that are also ROSAT All-Sky Survey stars 
(Alcal\'a et al. 1995), medium resolution spectra of all cluster
members (Mamajek et al. 1999, Lawson et al. 2002) and optical 
photometric study (Lawson et al. 2001, 2002).  For stars common 
to 2 or 3 studies, comparison of the spectral types suggests a 
typical uncertainty of $\pm 1$ subtype, a value comparable to 
the uncertainties present in the individual studies.  

In Figure 3, we find a strong correlation between the H$\alpha$ 
$EW$ and the ($K-L$) excess for the late-type members.  Adopting 
the same colour excess as in Figure 2, we again count the same
$7 \pm 2$ stars as having an IR excess when photometric errors 
are considered.  

\subsection{The disc fraction -- dust discs and accretion discs}

Merging the above results, we conclude $9 \pm 2$ out of the 
15 known stellar systems in the cluster, or a fraction of 
$0.60 \pm 0.13$ (2$\sigma$) have IR excesses.  This number also 
defines the fraction of stars with {\it dust discs\,}, or the 
{\it disc fraction\,} of the population.  However, a more
important parameter in PMS star evolution is the fraction of 
stars possessing {\it accretion discs\,} in a young stellar 
population.  By combining studies of populations of different 
ages it is possible to determine the timescale for the end of 
significant star-disc activity and probably also the timescale 
for Jovian planet building (e.g. Hillenbrand \& Meyer 1999).

\setcounter{figure}{1}
\begin{figure}
\begin{center}
\includegraphics[height=85mm]{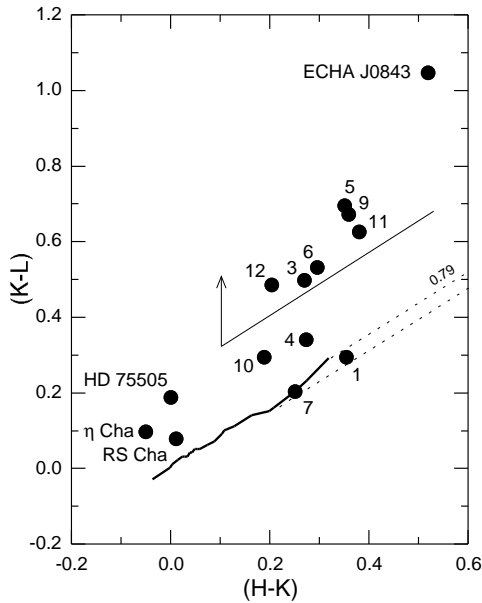}
\caption{($K-L$)/($H-K$) colour-colour diagram for
members of the $\eta$ Cha cluster surveyed at 3.5 $\mu$m. 
See the caption to Figure 1 for the meaning of the solid and 
dashed lines and other symbols.}
\end{center}
\end{figure}

In other recent studies the number of accreting stars 
has been determined from the ($H-K$) colour excess (e.g. 
Rebull et al. 2002). This has been a useful technique since 
the ($H-K$) colour is sensitive to warm inner discs, and also 
since {\it JHK\,} detector arrays have allowed entire
star formation regions to be surveyed.  With the introduction 
of $L$-band imagers, sensitive to cooler dust and to 
lower luminosity discs due to the increased contrast provided
by the $L$-band photometry,
we expect the disc fraction to increase as we 
demonstrate in Figure 1. But with the availability of $L$-band 
data, do we continue to detect accretion discs or are the 
$L$-band measurements increasingly sensitive to remnant dust 
discs in systems where there is no longer star-disc coupling?   

The ($K-L$) excesses measured for the $\eta$ Cha cluster stars 
(Figure 3) suggest a continuum of behaviour; from a CTT star 
(ECHA J0843.3--7905) showing strong star-disc 
interaction, stars that are CTT -- WTT transition objects still 
showing evidence for on-going accretion (RECX 5, 9 and 11), WTT 
stars with weak IR excesses (RECX 3, 6 and 12), and WTT stars 
with little or no IR excess (RECX 1, 4, 7 and 10, and ECHA 
J0841.5-7853).  Of the 15 known cluster members, the 4 stars 
with IR excesses $\Delta(K-L) > 0.4$ are all likely to 
be experiencing on-going accretion (the uncertainty in this 
number is $\pm 2$ stars; 2$\sigma$). We have spectroscopically 
confirmed that accretion is present in two cases: ECHA J0843.3-7905 
(Lawson et al.\ 2002) and RECX 11 (see below).  For most of
the late-type stars in the cluster, $\Delta(K-L) \approx 3 
\Delta(H-K)$, so our accretion criterion is little different 
from that adopted by Rebull et al. (2002), where $\Delta(H-K) 
> 0.15$.  These stars allow us to define an accretion 
fraction for the cluster of $0.27 \pm 0.13$ ($2\sigma$).  We 
further consider the issue of accretion in Section 4. 

In Figure 4 we show spectral energy distributions for 4 of the 
cluster stars.  We plot the measured fluxes of the CTT star ECHA 
J0843.3-7905, and offset the fluxes of the other stars to 
illustrate the range of IR signatures that are present in these 
stars.  ECHA J0843.3-7905 shows a flat spectrum at near-IR 
wavelengths with high colour excess.  On-going accretion in 
this star is supported by its rich optical emission spectrum, 
with a H$\alpha$ $EW = -110$ \AA.  The star is likely 
associated with IRAS F08450--7854.  The {\it IRAS\,} Faint 
Source Catalogue ({\it IRAS\,} FSC) indicates high-quality 
25-$\mu$m and 60-$\mu$m fluxes of 0.30 Jy and 0.28 Jy, 
respectively; approximately twice the $L$-band flux.  (The 
{\it IRAS\,} FSC indicates an upper limit 12-$\mu$m flux 
of 0.27 Jy.)  RECX 11 is one of 3 RECX stars with $\Delta(K-L) 
\approx 0.5$, and with a H$\alpha$ $EW \approx 7-20$ \AA\, that 
places them near the traditional CTT-WTT star `boundary' of 
activity.  MSSSO 1.9-m/coud\'e spectroscopy of RECX 11 shows 
the star is still accreting from its circumstellar disc, with 
broad (width $\sim 600$ km\,s$^{-1}$) and variable infall 
signatures at H$\alpha$ that are phased to the 3.95-d rotation 
period of the star (Lyo et al., in preparation).  RECX 11 
appears associated with IRAS F08487--7848, with 12-$\mu$m, 
25-$\mu$m and 60-$\mu$m fluxes of 0.29 Jy, 0.32 Jy and 0.27 Jy,
respectively.  RECX 6 is representative of WTT stars in the 
cluster with IR excesses of $\Delta(K-L) \approx 0.3$, and 
RECX 10 is a WTT star with little or no IR excess (see Figure 3).

\setcounter{figure}{2}
\begin{figure}
\begin{center}
\includegraphics[width=84mm]{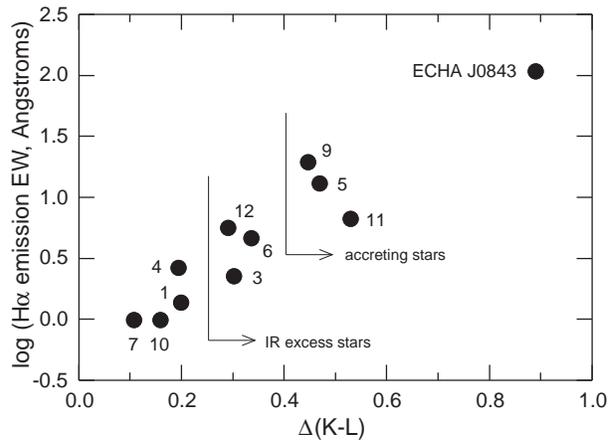}
\caption{H$\alpha$ emission-line {\it EW\,} versus 
$(K-L)$ excess for the late-type stars in the $\eta$ Cha 
cluster.  The individual stars are identified by their RECX 
number, except for the CTT star ECHA J0843.3--7905.  Stars 
with $\Delta(K-L) > 0.25$ are identified as IR excess 
objects with circumstellar discs.  Those stars with 
$\Delta(K-L) > 0.4$ are identified as stars with 
accretion discs, see Section 3.3.}
\end{center}
\end{figure}

\subsection{Additional considerations}

\subsubsection{Binarity}

In addition to the A7+A8 dual-lined eclipsing binary and $\delta$ 
Scuti system RS Cha AB, on-going study of the cluster
population has found several confirmed or probable binary systems.
Mamajek et al. (1999) noted that the H$\alpha$ emission profiles 
of RECX 7 and 9 were double, indicating that these stars may be 
spectroscopic binaries.  Lawson et al. (2001) noted several of 
RECX stars had elevated $V$ magnitudes compared to other cluster 
members of similar spectral type; RECX 9 and 12 are elevated by 
$\approx 0.7$ mag (suggesting near-equal mass systems) and 2 of 
the K-type stars (RECX 1 and 7) are $0.3-0.5$ mag brighter than 
the third K-type cluster member RECX 11.  

Speckle $K$-band imaging of RECX 1 and 9 by K\"ohler (2001)
found both stars have companions at separations of $0.1-0.2$
arcsec.  Observations of RECX 1 made during 1996 and 2000 showed
motion that might indicate a decade-long orbit.  Consideration 
of the stellar background density suggests these nearby stars 
are likely to be physically related to the primaries. If future
observations confirm these systems, RECX 1AB and RECX 9AB have 
$K$-band brightness ratios of $\approx 0.8$ and 0.5, respectively.
Observations of RECX 7 made by us using the 1.9-m telescope and 
coud\'e spectrograph at MSSSO during 2002 February showed RECX 7
to be a dual-lined spectroscopic 
binary with a period of 2.6 d (the same as the photometric 
period; Lawson et al. 2001) and a mass ratio of $\approx$ 2.3:1.
(RECX 7 has also been observed to be a spectroscopy binary 
by Donati, private communication.)  If RECX 7A is a near-solar 
mass star, then RECX 7B is a $\sim 0.4$ M$_{\odot}$ early-M star.
RECX 12 remains a candidate binary.  However, Lawson et al. 
(2001) found 2 periods (1.3 and 8.6 d) in the $V$-band light 
curve of the star in observations made in 1999 and 2000.  One 
of these periods may be the binary period.  

\setcounter{figure}{3}
\begin{figure}
\begin{center}
\includegraphics[height=85mm]{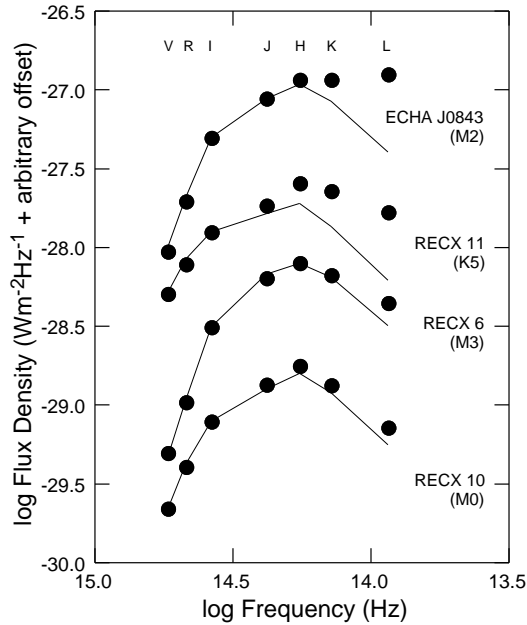}
\caption{Spectral energy distributions for ECHA J0843.3-7905, 
RECX 11, RECX 6 and RECX 10.  The solid lines represent spectral 
energy distributions for main-sequence stars of similar spectral
type to the cluster stars, derived from colours given by Kenyon 
\& Hartmann (1995), normalized to the $I$-band fluxes and assuming
no reddening.  The adopted spectral types are uncertain by $\pm 1$ 
subtype; see Section 3.2.  The worst-case photometric errors 
(the $5-10$ percent uncertainties in the $L$-band magnitudes)
translate to roughly the plotted size of the points.  The broad
band magnitudes were converted to fluxes using the conversions
of Bessell (1979), Bessell \& Brett (1988) and Koornneef (1983).}
\end{center}
\end{figure}

We have considered the effect of a binary companion on the IR
excesses determined from our {\it JHKL\,} photometry.  For a 
system such as RECX 7 with a mass ratio of $2-3$:1 we calculate, 
using the PMS models of Siess et al. (2000), that the near-IR 
colours will appear redder due to the presence of the secondary
by $< 0.03$ mag.  Binaries with higher mass ratios show less 
distortion of the primary star colours.  Due to the luminosity
ratio of such systems, any IR excess present is likely 
associated with the primary.  For distant near-equal mass 
systems such as RECX 1 the available data does not allow us to 
determine which member of the binary contains the IR excess, or 
if the excess is shared.  For close binaries, we would envisage 
a circumbinary disc if one were present; the 2 confirmed 
spectroscopic binaries in the cluster (RS Cha and RECX 7) have 
no IR excess (see Figure 2).  Since our criterion for 
the presence of an IR excess is $\Delta(K-L) > 0.25$, we conclude
that binarity will not significantly distort our results.

\subsubsection{Reddening}

Photometry of the cluster members indicates reddening is absent 
or low.  For the early-type stars, Westin (1985) found $E$($b-y$) 
$= -0.004$ for $\eta$ Cha, and the light curve of the binary RS Cha 
has been modelled assuming $E$($B-V$) = 0.0 (e.g., Clausen \& 
Nordstr\"om 1978).  For the late-type stars, our IR photometry 
indicates $\Delta(J-H) \approx 0.3\Delta(K-L)$, whereas the 
reddening vector in this plane has a gradient of 2.47 (see Figure 
1b).  We conclude that reddening is unimportant in our determination 
of the disc fraction.

\subsubsection{K-band excesses in the K-type stars}

Comparison of the IR colour excesses indicates the 3 K-type 
stars (RECX 1, 7 and 11) have $K$-band excesses of $\sim 0.1$ 
mag, which has the effect of increasing the ($H-K$) colours 
and decreasing the ($K-L$) colours.  This effect is not seen 
in the M-type stars, and cannot be attributed to reddening.  
Without 2-$\mu$m spectroscopy we cannot confirm if the excess 
is due to $K$-band activity, e.g. Greene \& Lada (1996)
detected Br$\gamma$ emission in half the CTT stars in a sample
of $\rho$ Oph PMS stars, although in none of the WTT stars.  
For RECX 1, an increase in the ($K-L$) colour of $\sim 0.1$ 
mag would make the IR excess significant.  However, the star 
is already included in our estimate of IR excess objects once 
we account for observational uncertainties.

\section{Summary and Conclusions}

Considerable uncertainty reigns concerning the longevity, or more
likely, the distribution of longevities, of circumstellar discs (Section
1).  Despite its modest population, the $\eta$ Cha cluster provides a
rare opportunity to examine -- at high sensitivity -- disc properties 
of PMS stars with intermediate ages whose selection is unbiased with 
respect to disc existence\footnote{The importance of the $\eta$ Cha
cluster in this respect is recognized by its inclusion in the first 
year guaranteed time programme of the {\it Space Infrared Telescope 
Facility}.}.  If discs decay rapidly as indicated by some past 
studies, then no discs at all are expected in this cluster.

We find, however, that 9/15 or 60 percent of $\eta$ Cha primaries 
show IR excesses in the $(H-K)/(K-L)$ diagram (the late-type stars 
identified in Figure 2, plus $\eta$ Cha and HD 75505).  The excess 
cannot be attributed to errors in photometry (Section 2), binarity 
(Section 3.4.1) or reddening (Section 3.4.2).  Long-lived circumstellar 
discs are the only plausible explanation.  One of these stars, ECHA
J0843.3-7905, is a CTT star with active accretion (Lawson et al.\ 2002),
and the H$\alpha - \Delta(K-L)$ correlation seen in the late-type 
population suggests that up to 3 other stars may be accreting (see
Figure 3).  High-resolution spectroscopic study now underway will
address this issue.

Why do we find a high disc fraction at $t \simeq 9$ Myr when some
other studies find discs largely disappear by $t = 3-6$ Myr?  We first
recognize that, except for ECHA J0843.3-7905, the $\eta$ Cha discs 
would have been mostly missed from {\it JHK\,} colours alone (see
Figure 1a).  Sensitive $L-$band surveys are essential for the detection
of aging PMS discs.  The principal discrepancy among $L-$band studies 
lies between our high disc fraction (9/15 or 0.60) in $\eta$ Cha and 
the low disc fraction (9/75 or 0.12) for the 5 Myr old cluster NGC
2362 found by Haisch et al. (2001).  We suggest several explanations
for this difference.  First, the assigned age of NGC 2362 PMS stars
relies solely on the turnoff age of the O9Ib supergiant $\tau$ CMa 
(Balona \& Laney 1996) and the assumption that all stars in the cluster 
are coeval.  Secondly, NGC 2362 is 1480 pc distant compared to 97 pc 
for $\eta$ Cha.  The distance ratio alone degrades the $L$-band 
sensitivity by a factor of 200.  Because of this, faint discs in 
NGC 2362 may not have been detected.  Also, the limiting mass of 
the Haisch et al. (2001) study of NGC 2362 is $M \approx 1$ M$_{\odot}$
(spectral type mid-K) compared to $M \approx 0.2$ M$_{\odot}$ (spectral 
type M4) in our study of $\eta$ Cha.  As we discuss in Section 3.1, 
it is easier to detect an IR excess in a late-M star, compared to 
a K-type star, using standard IR colour-colour plane analysis. Also, 
it is possible that disc lifetimes are shorter in higher mass stars.  
Third, there might be variance in the disc destruction rate amongst 
clusters due to different rates of close encounters or photoevaporation 
due to massive stars, e.g. the discs in NGC 2362 might have been 
stripped by the UV/wind of $\tau$ CMa (see Hollenbach et al. 2000 
for theory). 

A combination of the above factors can explain why there is a 
$2-4$ dispersion in the observed disc fraction for PMS star clusters 
of a similar age (Hillenbrand \& Meyer 1999).  While noting these 
differences, our results seen together with studies of other older 
nearby PMS stars (e.g. study of the TW Hydrae Association members 
by Muzerolle et al. 2000) indicate that IR-detected discs can be
present in $\sim 60$ percent, and accretion discs can be present
in $\sim 30$ percent, of $\sim 10$ Myr-old PMS stars.  

\section*{Acknowledgments}

We thank C. Kaminski for obtaining our SPIREX/Abu observations, 
and J. Kastner and colleagues at the Rochester Institute of 
Technology for performing the pipeline analysis of the raw 
observations.  We also thank M. Burton (UNSW) for coordinating 
the Australian allocation of SPIREX observing time, S. James
for installing the SPIREX/Abu reduction package at ADFA, and 
N. Zarate (NOAO) at the IRAF help desk for on-line assistance.  
SPIREX was a facility operated by the Center for Astrophysical
Research in Antartica.  We thank the MSSSO and SAAO time allocation 
committees for telescope time during 2002 which enabled us to 
complete this project.  MSSSO is operated by the Research School 
of Astronomy and Astrophysics, Australian National University.  
SAAO is a national facility operated by the National Research 
Foundation of South Africa.  ARL acknowledges the support of a 
UNSW/ADFA International Postgraduate Research Scholarship.  WAL 
acknowledges support from the UNSW Research Support Programme 
and UNSW/ADFA Special Research Grants.  EEM thanks the SIRTF 
Legacy Science Program for support.  EDF's research is supported 
in-part by NASA contracts NAS8-38252 and NAG5-8422.  LAC is 
supported by a National Research Foundation Post-graduate 
Scholarship.

\bsp

\label{lastpage}

\end{document}